\documentclass[aps,prc,twocolumn,superscriptaddress]{revtex4}  

\usepackage{mathrsfs,amsmath,amssymb,amsthm,natbib}
\usepackage{amssymb,fmtcount}
\usepackage{graphicx,epsfig,latexsym,overpic,amssymb,color}
\usepackage{threeparttable}
\usepackage{qcircuit}
\usepackage{multirow}
\usepackage{siunitx}
\usepackage{makecell} 
\preprint{\today}

\newcommand{\cn} {\sc{cnot}}

\begin{document}
\title{
Quantum computing of the pairing Hamiltonian at finite temperatures
}
\author{Chongji Jiang}
\affiliation{
State Key Laboratory of Nuclear Physics and Technology, School of Physics,
Peking University, Beijing 100871, China
}
\author{Junchen Pei}\email{peij@pku.edu.cn}
\affiliation{
State Key Laboratory of Nuclear Physics and Technology, School of Physics,
Peking University, Beijing 100871, China
}
\affiliation{
Southern Center for Nuclear-Science Theory (SCNT), Institute of Modern Physics, Chinese Academy of Sciences, Huizhou 516000,  China
}

\begin{abstract}
In this work, we study the pairing Hamiltonian with four particles at finite temperatures
on a quantum simulator and a superconducting quantum computer. The excited states are obtained
by the variational quantum deflation (VQD). The error-mitigation methods are  applied to improve the noisy results.
 The simulation of thermal excitation states is performed using the same variational circuit as at zero temperature. 
 The results from quantum computing become close to exact solutions at high temperatures, and demonstrate a smooth superfluid-normal phase transition  as a function of
temperatures as expected in finite systems.

\end{abstract}

\maketitle

\section*{\uppercase\expandafter{\romannumeral1}. Introduction}

The simulation of quantum many-body systems on quantum computers has
natural advantages by avoiding the exponential scaling of computing costs on classical computers~\cite{abrams}.
Atomic nuclei are strongly correlated finite quantum many-body systems, for which
the accurate treatment of many-body correlations is essential.
There are already several applications of quantum computing in nuclear physics,
such as the implementation of coupled cluster method for light nuclei~\cite{cloud}, the Lipkin model~\cite{lipkin1,lipkin2}, neutrino-nucleus scattering~\cite{neutrino},
nuclear dynamics~\cite{papenbrock,weijie}, and the symmetry restoration~\cite{lacrox} on quantum computers.
Presently these applications in simplified many-body  models
paved a route to practical quantum computing of small quantum systems such as light nuclei
in the near future.

Actually small quantum systems has novel features compared to large systems.
For large systems the statistical methods or mean-field theories are often suitable theoretical tools.
In particular, there is a superfluid-normal phase transition in large systems with increasing
temperatures but the phase transition is absent in finite small systems.
In this respect, the finite-temperature BCS or Hartree-Fock-Bogoliubov theory is breakdown which
results in a false pairing phase transition in nuclei~\cite{Goodman1981}.
With elaborate many-body approaches, such as the quantum monte-carlo~\cite{alhassid}
and particle number projections at finite temperatures~\cite{Esebbag}, the false phase transition
is washed out. In addition, the existence of a pseudogap phase
 in high-$T_c$ superconductors has been widely studied  and the origin of the pseudogap remains an open question~\cite{pseudo,pseudo2}.
Indeed, the exact treatment of thermal excitations of quantum systems has broad implications in static and dynamical observables.
The accurate descriptions of hot nuclei and nuclear matter are relevant for descriptions of
 level densities~\cite{fanto}, shape transitions~\cite{alhassid,martin},  fission barriers~\cite{pei09} and equation of state for neutron stars~\cite{eos}.

There are several quantum algorithms to simulate many-body systems on quantum computers.
The widely used variational quantum eigensolver (VQE) is a robust and flexible way to compute the ground
state of a Hamiltonian~\cite{cloud,VQE}. The modified VQE, namely variational quantum deflation (VQD),  can be applied to excited states~\cite{effh}.
In addition, the quantum phase estimation method can also solve the eigenstate problems
which requires deep circuits with ancilla qubits~\cite{lacrox}.
On the other hand, the hybrid quantum and classical computing has been extensively studied
so that the optimization of VQE is feasible~\cite{hybrid}. Besides quantum algorithms, the development of quantum computing hardware is fast and IBM is expected to
deliver a 4000-qubit system by 2025.

In this work, the ground state, excited states, and thermal states of the pairing
Hamiltonian are studied with the variational quantum computation.
One of the key issues is to apply VQD to solve excited states and their degeneracies.
Although the quantum computing of the pairing Hamiltonian and the Lipkin model have been studied in the literatures~\cite{lipkin1,lipkin2,lacrox},
a comprehensive study of eigenstates and thermal states is still inspiring.
The finite-temperature BCS results with a false phase transition are also shown for comparison
to emphasize the significance of quantum computing. The calculations are firstly performed with the simulator
$\textsl{Qiskit}$~\cite{qiskit}. Then practical quantum computing is performed on a superconducting quantum computer provided by IBM.
The error mitigation methods for the noisy quantum computing have also been discussed.

\section*{\uppercase\expandafter{\romannumeral2}. Theoretical Framework}

This work solves the pairing Hamiltonian in a degenerate shell space, which has exact solutions for benchmark of different methods.
It is known that the fermionic operators can be implemented on quantum computers with the Jordan-Wigner transformation~\cite{cloud,lacrox,JW, JW2}.
For the pairing Hamiltonian, it is more efficient to map the pairs with the quasi-spin operators~\cite{JW3,lacrox2}. 
In this work, the pairing Hamiltonian is rewritten with the quasi-spin operators as~\cite{manyb}:
\begin{subequations}
\begin{equation}
H=-G\sum\limits_{m,m'>0}a_{m'}^{+}a_{-m'}^{+}a_{-m}a_{m}=-G\sum\limits_{m,m'>0} s_{+}^{(m')}s_{-}^{(m)}
\end{equation}
with
\begin{equation} s_{+}^{(m)}=a_{m}^{+}a_{-m}^{+},\quad s_{-}^{(m)}=a_{-m}a_{m}.
\end{equation}
\end{subequations}
where two orbitals of $(m, -m)$  form a pair. 
The quasi-spin operators $s_{+}$, $s_{-}$ have the same commutation properties of angular momentum operators.

It is convenient to map the transformed pairing Hamiltonian in the quasi-spin basis into qubit basis:
\begin{subequations}
\begin{equation}
H = -G\sum\limits_{p,q>0} s_{+}^{(p)}s_{-}^{(q)},
\end{equation}
where the operators can be represented by Pauli matrices:
\begin{equation}
s_{+}^{(p)}s_{-}^{(p)}\rightarrow\frac{1}{2}(I^{(p)}+\sigma_{z}^{(p)}),
\end{equation}
\begin{equation}
s_{+}^{(p)}s_{-}^{(q)}\rightarrow\frac{1}{2}(\sigma_{x}^{(p)}\otimes\sigma_{x}^{(q)}+\sigma_{y}^{(p)}\otimes\sigma_{y}^{(q)}).
\end{equation}
\label{espon}
\end{subequations}

    \subsection{Details of the pairing Hamiltonian}

In the following part, we describe the pairing Hamiltonian  being mapped into the qubit basis.
We study $N$=4 particles in a (2$j$+1)-fold degenerate $j$-shell corresponding to $\Omega=3$ and  $\Omega=4$, where $\Omega$=$j+\frac{1}{2}$ is
the number of pairs.
In the shell model, the configuration spaces for $\Omega=3$ and $\Omega=4$ are $C_6^4=15$ dimensional and  $C_8^4=70$  dimensional, respectively.
The complexity of classical computing  increases exponentially with the configuration space $\Omega$.
The half-occupied configuration space leads to the largest computing costs.
We will show that the pairing Hamiltonian can be simulated
with 3 qubits for $\Omega=3$ and with 4 qubits for $\Omega=4$ and so on, which is irrespective of the number of particles.

For the case of  $\Omega=3$,  the pairing Hamiltonian can be constructed on 3 qubits.
The transformed pairing Hamiltonian is represented in terms of pauli matrices according to Eq.\ref{espon}.
The pairing Hamiltonian of $\Omega=3$ can be solved in the qubit basis space of
$\left\{\left|\uparrow\uparrow\uparrow\right\rangle,\left|\uparrow\uparrow\downarrow\right\rangle,
\left|\uparrow\downarrow\uparrow\right\rangle,\left|\uparrow\downarrow\downarrow\right\rangle,\left|\downarrow\uparrow\uparrow\right\rangle,
\left|\downarrow\uparrow\downarrow\right\rangle,\left|\downarrow\downarrow\uparrow\right\rangle,\left|\downarrow\downarrow\downarrow\right\rangle\right\}$.
For $N$=4 and $\Omega=3$, the eigenspace can be reduced to $\left\{\left|\uparrow\uparrow\downarrow\right\rangle,
\left|\uparrow\downarrow\uparrow\right\rangle,\left|\downarrow\uparrow\uparrow\right\rangle\right\}$ 
since the number of particles is related to the $z$-component of total spin. 
For  $\Omega=4$ and $N$=4, the pairing Hamiltonian
can be represented on 4-qubits in a reduced eigenspace of ({$\left|\uparrow\uparrow\downarrow\downarrow\right\rangle$,$\left|\uparrow\downarrow\uparrow\downarrow\right\rangle$,$\left|\downarrow\uparrow\uparrow\downarrow\right\rangle$, $\left|\uparrow\downarrow\downarrow\uparrow\right\rangle$,$\left|\downarrow\uparrow\downarrow\uparrow\right\rangle$,$\left|\downarrow\downarrow\uparrow\uparrow\right\rangle$}).
The exact solutions of the pairing Hamiltonian of $\Omega=3$ and $\Omega=4$ with 4 particles are  given in Table I.

\renewcommand\arraystretch{1.5}
\begin{table*}[thb]
\caption{\label{tab:table1} The exact solution in the qubit basis space for the pairing Hamiltonian with $N$=4 particles in a degenerate shell space of $\Omega=3$ and $\Omega=4$.
In the table, $S_0$ denotes the $z$-component of the total spin and $s$ denotes the seniority number based on the seniority model. }
\centering
\begin{tabular}{|p{1.0cm}<{\centering}|p{0.5cm}<{\centering}|p{4cm}<{\centering}|p{5.5cm}<{\centering}|p{0.5cm}<{\centering}|p{2cm}<{\centering}|}
\hline
$(\Omega,N)$ & $S_0$  & Basis & Eigenstates & $s$ & Eigenvalue \\
\hline
\multirow{3}{*}{$(3,4)$} & \multirow{3}{*}{$\dfrac{1}{2}$} & \multirow{3}{*}{\makecell[c]{$\left|\uparrow\uparrow\downarrow\right\rangle$,$\left|\uparrow\downarrow\uparrow\right\rangle$,$\left|\downarrow\uparrow\uparrow\right\rangle$}} & $\frac{1}{\sqrt{3}}(\left|\uparrow\uparrow\downarrow\right\rangle+\left|\uparrow\downarrow\uparrow\right\rangle+\left|\downarrow\uparrow\uparrow\right\rangle)$   & $0$ & $-4G$ \\ \cline{4-6}
  &   &  & $\frac{1}{\sqrt{2}}(\left|\uparrow\uparrow\downarrow\right\rangle-\left|\uparrow\downarrow\uparrow\right\rangle)$ & \multirow{2}{*}{$2$} & \multirow{2}{*}{$-G$} \\ \cline{4-4}
  &   &  & $\frac{1}{\sqrt{6}}(-\left|\uparrow\uparrow\downarrow\right\rangle-\left|\uparrow\downarrow\uparrow\right\rangle+2\left|\downarrow\uparrow\uparrow\right\rangle)$  & &  \\ \hline
  \multirow{6}{*}{$(4,4)$} & \multirow{6}{*}{$0$} & \multirow{6}{*}{\makecell[c]{$\left|\uparrow\uparrow\downarrow\downarrow\right\rangle$,$\left|\uparrow\downarrow\uparrow\downarrow\right\rangle$,$\left|\downarrow\uparrow\uparrow\downarrow\right\rangle$,\\  $\left|\uparrow\downarrow\downarrow\uparrow\right\rangle$,$\left|\downarrow\uparrow\downarrow\uparrow\right\rangle$,$\left|\downarrow\downarrow\uparrow\uparrow\right\rangle$}} & $\frac{1}{\sqrt{6}}(\left|\uparrow\uparrow\downarrow\downarrow\right\rangle+\left|\uparrow\downarrow\uparrow\downarrow\right\rangle+\left|\downarrow\uparrow\uparrow\downarrow\right\rangle+\left|\uparrow\downarrow\downarrow\uparrow\right\rangle+\left|\downarrow\uparrow\downarrow\uparrow\right\rangle+\left|\downarrow\downarrow\uparrow\uparrow\right\rangle)$  & $0$ & $-6G$  \\ \cline{4-6}
   &   &   & $\frac{1}{\sqrt{2}}(\left|\downarrow\uparrow\uparrow\downarrow\right\rangle-\left|\uparrow\downarrow\downarrow\uparrow\right\rangle)$ & \multirow{3}{*}{$2$} & \multirow{3}{*}{$-2G$} \\ \cline{4-4}
  &    &   & $\frac{1}{\sqrt{2}}(\left|\downarrow\downarrow\uparrow\uparrow\right\rangle-\left|\uparrow\uparrow\downarrow\downarrow\right\rangle)$  & & \\ \cline{4-4}
   &   &    & $\frac{1}{\sqrt{2}}(\left|\downarrow\uparrow\downarrow\uparrow\right\rangle-\left|\uparrow\downarrow\uparrow\downarrow\right\rangle)$  & & \\ \cline{4-6}
  &    &  & $\frac{1}{2}(\left|\uparrow\downarrow\uparrow\downarrow\right\rangle-\left|\downarrow\uparrow\uparrow\downarrow\right\rangle-\left|\uparrow\downarrow\downarrow\uparrow\right\rangle+\left|\downarrow\uparrow\downarrow\uparrow\right\rangle)$ & \multirow{2}{*}{$4$} & \multirow{2}{*}{$0$} \\ \cline{4-4}
   &   &  & $\frac{\sqrt{3}}{3}(\left|\uparrow\uparrow\downarrow\downarrow\right\rangle+\left|\downarrow\downarrow\uparrow\uparrow\right\rangle)-\frac{\sqrt{3}}{6}(\left|\downarrow\uparrow\uparrow\downarrow\right\rangle+\left|\uparrow\downarrow\downarrow\uparrow\right\rangle+\left|\uparrow\downarrow\uparrow\downarrow\right\rangle+\left|\downarrow\uparrow\downarrow\uparrow\right\rangle)$  & & \\
\hline
\end{tabular}
\end{table*}

\subsection{State preparation}

Next we prepare the trial state on the quantum circuits.
To simplify the quantum circuits, the symmetry of particle number conservation is exploited.
For $\Omega=3$ with $N$=4 particles,
the ansatz wave function with 2 variational parameters is represented as:
\begin{equation}
\vert\psi\rangle_{t}=\sin{\frac{\theta}{2}}\left|\uparrow\uparrow\downarrow\right\rangle+\cos{\frac{\theta}{2}}\sin\eta\left|\uparrow\downarrow\uparrow\right\rangle+\cos{\frac{\theta}{2}}\cos\eta\left|\downarrow\uparrow\uparrow\right\rangle
\end{equation}
The quantum circuit of $\Omega=3$ is shown in Fig.1. The rotation angles $\eta,\theta$
correspond to the variational parameters.

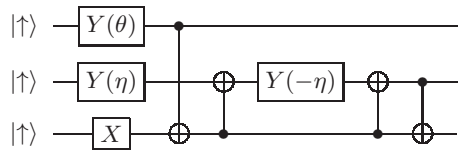
\begin{figure}[hb]
\begin{center}
\[ \Qcircuit @C=1em @R=.7em {
\lstick{\left\vert\uparrow\rangle\right.} & \gate{Y(\theta)} & \ctrl{2} & \qw & \qw & \qw & \qw & \qw \\
\lstick{\left\vert\uparrow\rangle\right.} & \gate{Y(\eta)} & \qw & \targ & \gate{Y(-\eta)} & \targ & \ctrl{1} & \qw \\
\lstick{\left\vert\uparrow\rangle\right.} & \gate{X} & \targ & \ctrl{-1} & \qw & \ctrl{-1} & \targ & \qw
}
\]
\caption{Quantum variational circuit for $\Omega=3$ and $N$=4 on 3 qubits, in which the $Y(\theta)$  gate
performs a rotation of $\theta$ around the $Y$-axis direction.}
\end{center}
\end{figure}

For $\Omega=4$ with 4 particles, the trial wave function in computational basis can be expressed by variational rotation angles $\theta_0,...,\theta_4$.
The associated quantum circuit with 4-qubits are shown in Fig.2.  For $\Omega=4$ with 6 particles, the circuit can be much simpler with a reduced eigenspace.
The circuit for $\Omega=4$ is constructed according to the ansatz that preserves the number of particles, as presented in Refs.~\cite{6Li,EPA1,EPA2}.
This circuit employs 5 two-qubit building blocks as shown in Fig.2. Each building block $U(\theta)$ is written as,
\begin{equation}
    U(\theta) = \left(
    \begin{matrix}
    1 & 0 & 0 & 0 \\
    0 & \sin\theta & \cos\theta & 0 \\
    0 & \cos\theta & -\sin\theta & 0 \\
    0 & 0 & 0 & 1 \\
    \end{matrix}\right)
\end{equation}
which is a unitary variational  transformation but preserves the number of particles.
In this way, the circuit is rather low-depth and results from quantum computing are less noisy.
The circuits for systems with a larger $\Omega$ can also be constructed efficiently using the two-qubit building blocks.

\begin{figure*}[bt]
\begin{center}
\[ \Qcircuit @C=.45em @R=.8em {
\lstick{\left\vert\uparrow\rangle\right.} & \qw & \qw & \targ & \qw & \ctrl{1} & \qw & \targ & \qw & \qw & \qw & \qw & \qw & \qw & \qw & \qw & \qw & \targ & \qw & \ctrl{2} & \qw & \targ & \qw & \qw & \qw & \qw & \qw & \qw & \qw & \qw  \\
\lstick{\left\vert\uparrow\rangle\right.} & \gate{X} & \qw & \ctrl{-1} & \gate{Y(\theta_0)} & \targ & \gate{Y(-\theta_0)} & \ctrl{-1} & \qw & \qw & \ctrl{2} & \gate{Y(\theta_2)} & \targ & \gate{Y(-\theta_2)} & \ctrl{2} & \qw & \qw & \qw & \qw & \qw & \qw & \qw & \qw & \qw & \ctrl{1} & \gate{Y(\theta_4)} & \targ & \gate{Y(-\theta_4)} & \ctrl{1} & \qw   \\
\lstick{\left\vert\uparrow\rangle\right.} & \gate{X} & \qw & \ctrl{1} & \gate{Y(\theta_1)} & \targ & \gate{Y(-\theta_1)} & \ctrl{1} & \qw & \qw & \qw & \qw & \qw & \qw &\qw & \qw & \qw & \ctrl{-2} & \gate{Y(\theta_3)} & \targ & \gate{Y(-\theta_3)} & \ctrl{-2} & \qw & \qw & \targ & \qw & \ctrl{-1} & \qw & \targ & \qw \\
\lstick{\left\vert\uparrow\rangle\right.} & \qw & \qw & \targ & \qw & \ctrl{-1} & \qw & \targ & \qw & \qw & \targ & \qw & \ctrl{-2} & \qw & \targ & \qw & \qw & \qw & \qw & \qw & \qw & \qw & \qw & \qw & \qw & \qw & \qw & \qw & \qw & \qw
\gategroup{1}{4}{2}{8}{1.2em}{--}
\gategroup{3}{4}{4}{8}{1.2em}{--}
\gategroup{2}{11}{4}{16}{1.24em}{--}
\gategroup{1}{18}{3}{22}{1.2em}{--}
\gategroup{2}{25}{3}{29}{1.2em}{--}
}
\]
\caption{Quantum variational circuit for $\Omega=4$ and $N$=4 on 4 qubits, which includes 5 two-qubit building blocks.}
\end{center}
\end{figure*}
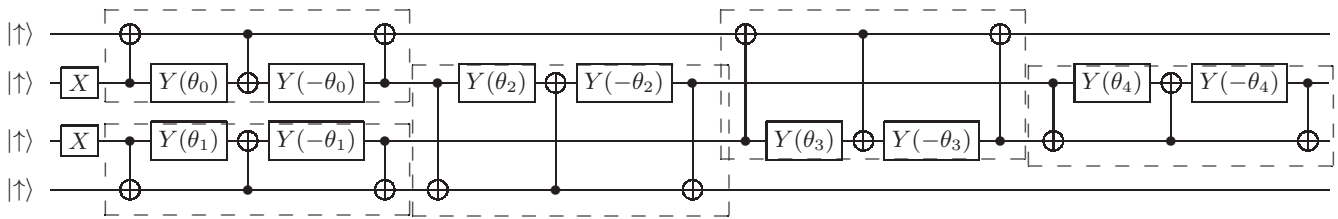

{\centering\subsection{VQD for Excited States}}

Firstly the ground state solution is obtained by adjusting the variational parameters in the Hamiltonian.
\begin{equation}
E(\lambda)\equiv\left\langle\psi(\lambda)\vert H\vert\psi(\lambda)\right\rangle
\end{equation}
The  ansatz state $\left\vert\psi(\lambda)\right\rangle$ is prepared with variational parameters $\lambda$,
which are the rotation angles in quantum gates of the circuit.
The ground state corresponds to the minimum energy by making measurements of Pauli terms of the pairing Hamiltonian.
For circuits with multiple parameters, the optimized numerical method is needed to search the minimum.

VQD is a modification of VQE and can be applied to compute excited states~\cite{effh}.
For the $k$-th excited state, the variational parameters $\lambda_k$ for the ansatz state $\psi(\lambda_k)$ are obtained by minimizing the extended cost function as
\begin{equation}
E(\lambda_k)\equiv\left\langle\psi(\lambda_k)\vert H\vert\psi(\lambda_k)\right\rangle + \sum_{i=0}^{k-1}\beta_i |\langle \psi(\lambda_k)\vert\psi(\lambda_i) \rangle |^2
\end{equation}
This means that $\psi(\lambda_k)$ is required to be orthogonal to lower states.
This is equivalent to solve  an effective Hamiltonian $H(k)$ for $k$-th excited states $\vert k\rangle$ as follow,
\begin{equation}
H(k)\equiv H+\sum_{j=0}^{k-1}\beta_j\left\vert j\rangle\langle j\right\vert.
\label{excit}
\end{equation}
Here $H$ represents pairing Hamiltonian, $\vert 0\rangle$ represents ground state and $\vert j\rangle$ represents $j$-th excited state.
The excited states are solved successively with increasing excitation energies.
The parameters $\beta_j$ are large values that shift lower states to higher energies.
The excited states and ground state share the same basis and the same circuit with different variational parameters.

To implement VQD, it is crucial to calculate the wave function overlap $\vert\langle j \vert k \rangle\vert^2$,
which is realized by $\vert\langle 0 \vert U^{\dagger}(j)U(k)\vert 0\rangle\vert^2$~\cite{effh}.
We can prepare the state $U^{\dagger}(j)U(k)\vert 0\rangle$ using the trial state circuit followed by the inverse
of the previously-computed state. The overlap is obtained by measuring the final probabilities of $\vert\uparrow\uparrow\uparrow\rangle$.
 This method requires the same number of qubits as VQE and at most twice the circuit depth.
Note that there are several methods to compute excited states such as the quantum phase estimate method~\cite{abrams,QPE2,lacrox}, the quantum Lanczos method~\cite{qLanczos,lacrox2}
and the quantum equation of motion~\cite{qEOM,lipkin2}.
 VQD works for general Hamiltonian problems and requires least resources to compute excited states compared to these methods.
In applications of the VQD method, one should be cautious since  errors would be accumulated and become larger for higher states.

\vspace{10pt}

\section*{\uppercase\expandafter{\romannumeral3}. Systems at Finite Temperatures }

{\centering \subsection*{A. Seniority Model}}

The exact eigenvalues of the degenerate pairing hamiltonian with $N$ particles can be obtained by the seniority model as~\cite{manyb}
\begin{equation}
    E_N^s =-\frac{G}{4}(N-s)(2\Omega-N-s+2),
\end{equation}
where $s$ is the seniority quantum number representing the number of unpaired nucleons.

The eigenstates are usually degenerate except for the ground state. The degeneracy of excited states is related to $s$ as~\cite{manyb}:
\begin{equation}
    d_s=\binom{\Omega}{s/2}-\binom{\Omega}{s/2-1},
\end{equation}
while $s$  satisfy $s\leq N$. It is consistent with the results of exact diagonalization of pairing Hamiltonian shown in Table I.

The system at a finite temperature $kT$ is described by the canonical ensemble.
The partition function can be written as
\begin{equation}
    Z_c=\sum_{\substack{s}}d_s\exp{[-\beta(E_N^s-E_N^0)]},
\end{equation}
where $\beta=1/kT$ and $k$ is the Boltzmann constant.
The pairing energy at the finite temperature $kT$ is given by
\begin{equation}
    \langle H\rangle=\frac{1}{Z_c}\sum_{s} d_s E_N^s\exp{[-\beta(E_N^s-E_N^0)]}
\end{equation}

\begin{table*}[htb]
 \caption{\label{table1} The solved eigenspace of the pairing Hamiltonian for $\Omega=3$. This table also lists the eigen energies of the ground state (gs), the  first  degenerate excited states (1st es) from the $\textsl{Qiskit}$ simulator,
 raw quantum computing energies from $\textsl{IBM\_oslo}$, the readout error-mitigated energies (R-Miti), the zero-noise extrapolation energies (ZNE), and  energies from combined readout error-mitigation
 and zero-noise extrapolation. See text for details.
 }
\begin{ruledtabular}
\begin{tabular}{ccccccc}
  Eigenstate & $\vert\psi\rangle$ & E($\textsl{Qiskit}$) & E($\textsl{IBM\_oslo}$) & E(R-Miti) & E(ZNE) & E(R-Miti+ZNE) \\ \hline
     gs& $0.588\left\vert\uparrow\uparrow\downarrow\right\rangle+0.554\left\vert\uparrow\downarrow\uparrow\right\rangle+0.590\left\vert\downarrow\uparrow\uparrow\right\rangle$ & -4.013 & -3.751 & -3.871 &  -3.919  &  -4.123 \\ \hline
    \multirow{2}{*}{1st es} & $0.309\left\vert\uparrow\uparrow\downarrow\right\rangle+0.651\left\vert\uparrow\downarrow\uparrow\right\rangle-0.693\left\vert\downarrow\uparrow\uparrow\right\rangle$ & -1.024 & -1.126 & -1.107  &  -1.093 &  -1.064 \\ 
     & $0.809\left\vert\uparrow\uparrow\downarrow\right\rangle-0.569\left\vert\uparrow\downarrow\uparrow\right\rangle-0.146\left\vert\downarrow\uparrow\uparrow\right\rangle$ & -1.004 & -1.109 &  -1.086 &  -1.006 &  -0.978
\end{tabular}
\end{ruledtabular}
\end{table*}

\begin{table}[tb]
 \caption{\label{table1} The eigen energies of the pairing Hamiltonian for $\Omega=4$ from practical quantum computing.
 The readout error-mitigation (R-Miti) and zero-noise extrapolation (ZNE) results are also shown.
 }
\begin{ruledtabular}
\begin{tabular}{ccccc}
  Eigenstate & E($\textsl{Qiskit}$) & E($\textsl{IBM\_oslo}$) &  E(R-Miti) & E(ZNE) \\ \hline
    gs & -5.958 & -4.534 & -4.754 & -5.376\\ \hline
    \multirow{3}{*}{1st  es} & -2.020 & -2.108 &-2.124 & -2.055\\
     & -2.000 &-1.963 &-1.976 & -1.895\\
     &  -2.004 & -2.071 &-2.097 &-2.107 \\ \hline
     \multirow{2}{*}{2nd es} &  -0.064 &-0.809 &-0.728 &-0.620\\
     &  -0.015 &-0.655 &-0.559 &-0.233
\end{tabular}
\end{ruledtabular}
\end{table}

{\centering\subsection*{B. Finite Temperature BCS Theory}}

The finite temperature BCS (FT-BCS) or Bogoliubov theory has been widely used for descriptions of compound nuclei~\cite{Goodman1981}.
The partition function is based on quasi-particle excitations.
With FT-BCS,  the pairing gap equation is written as~\cite{Goodman1981},
\begin{equation}
    \Delta=\Delta_0\tanh{(\frac{1}{2}\beta\Delta)},
\end{equation}
where $\Delta_0=\frac{G\Omega}{2}$ is the BCS gap at zero temperature.
The expectation value of the pairing Hamiltonian is
\begin{equation}
    \langle H\rangle_{{\rm FT-BCS}}=-G\frac{N^2}{4\Omega}-\frac{\Delta^2}{G}
    \label{ftbcs}
\end{equation}
Within the FT-BCS theory, there is a ``phase transition" from a paired state to a normal state
at a critical temperature corresponding to
$kT_c=\frac{1}{2}\Delta_0$~\cite{Goodman1981}. The critical temperature is around 0.7 MeV for compound nuclei~\cite{martin,khan}.

\par
\vspace{10pt}

{\centering \subsection*{C.  Thermal States by VQE}
}

Thermal excited states are  mixed states which can be described by the density matrix:
\begin{equation}
    \rho_\beta=\frac{1}{Z_c}\sum_i e^{-\beta E_i}\vert\psi_i\rangle\langle\psi_i\vert,
\end{equation}
note that $\psi_i$ is the eigenstate of Hamiltonian. The expectation value of an observable $\hat{O}$ is defined by $\langle\hat{O}\rangle=\mathrm{Tr}(\rho_\beta\hat{O})$.
The pairing energy can be calculated by:
\begin{equation}
    \langle H\rangle=\frac{1}{Z_c}\sum_i e^{-\beta E_i}\langle\psi_i\vert H\vert\psi_i\rangle \equiv \langle\psi_c\vert H\vert\psi_c\rangle,
\end{equation}
while $\vert\psi_c\rangle\equiv\dfrac{1}{\sqrt{Z_c}}\sum_i e^{-\beta E_i/2}\vert\psi_i\rangle$.
Actually $\vert\psi_c\rangle$ are not known as a prior and is supposed to be determined by VQE.
The preparation of thermal equilibrium states with unitary quantum operations
is not trivial. It is known that a mixed thermal state can be generated by thermofield double states~\cite{tfd},
however, it is difficult to be applied to a general Hamiltonian.
Besides, the quantum imaginary time evolution has only been applied to geometric local Hamiltonians for thermal states~\cite{imaginary}.
The quantum computing of zeros of the partition function is an alternative way to study phase transitions and thermodynamic quantities~\cite{pzero}.
In our case, it is possible to construct a superposition of eigenstates without off-diagonal elements
to simulate the mixed thermal states.

Based on VQE, the thermal state can be determined by  minimizing the free energy $F$:
\begin{equation}
    F = \langle H\rangle-TS.
\end{equation}
The first term $ \langle H\rangle$ is the easy to calculate based on the variational wave function.
In regard to a mixed state, which is described by $\rho=\sum_i p_i \vert \psi_i\rangle\langle \psi_i\vert$, the definition of its Von Neumann entropy is $-k\sum_i p_i\ln{p_i}$.
The probability $p_i$ is the overlap between $\vert \psi_i\rangle$ and $\vert \psi_c\rangle$.
The second term of free energy can be expressed as
\begin{equation}
    T\langle S\rangle=-kT\sum_i \vert\langle\psi_i\vert\psi_c\rangle\vert^2\ln{\vert\langle\psi_i\vert\psi_c\rangle\vert^2},
\end{equation}
while $\vert\psi_i\rangle$ are previously-computed eigenstates of the pairing Hamiltonian at zero temperature. Finally, the cost function can be written as
\begin{equation}
\begin{array}{ll}
    \langle\psi(\lambda)\vert F\vert\psi(\lambda)\rangle = & \langle\psi(\lambda)\vert H\vert\psi(\lambda)\rangle+\\
     & kT\sum_i \vert\langle\psi_i\vert\psi(\lambda)\rangle\vert^2\ln{\vert\langle\psi_i\vert\psi(\lambda)\rangle\vert^2}\\
\end{array}
\label{entropyc}
\end{equation}
while $\psi(\lambda)$ is the trial wave function. The quantum computing  of overlaps is described in the implementation of VQD.
The variational parameters are obtained by minimizing the free energy in case eigenvalues are not known or not precise.
The variational measurements of $<H>$ can be implemented on the same variational circuit even without accurate knowledge of  eigenfunctions.
This method would be attractive if the entropy can be measured more efficiently.

In practical calculations, the number of variational parameters can be reduced by considering
the degeneracy of excited states. For $\Omega=3$,  the eigenspace is 3-dimensional, as shown in Table I.
To describe the superposition state,  we need at least two parameters.
Considering the degeneracy, the parameter space can reduce to 1-dimensional, such as $\psi(\alpha)=\cos{\alpha}\vert0\rangle+\frac{\sin\alpha}{\sqrt{2}}(\vert1\rangle+\vert2\rangle)$.
The state preparation circuit for calculating excited states can also be used
 for thermal states, in which the two variational parameters satisfy
 \begin{subequations}
\begin{equation}
    \eta=-\frac{\pi}{4}+\arctan(\frac{2\cot{\alpha}-1}{\sqrt{3}})
\end{equation}
\begin{equation}
    \theta=2\arccos{(\frac{\cos{\alpha}+\sin{\alpha}}{\sqrt{3}})}
\end{equation}
\end{subequations}

For $\Omega=4$ and $N$=4, the eigenspace is 6 dimensional as shown in Table I and 5 variational parameters are needed.
By considering the degeneracy,
we can construct the trial wave function in a 2-dimensional parameter space $\{\alpha,\beta\}$, $\left\vert\psi\right\rangle = \cos\alpha\left\vert gs\right\rangle + \frac{\sin\alpha\cos\beta}{\sqrt{3}}\sum\limits_{i=0,1,2}\left\vert E_1\right\rangle_i + \frac{\sin\alpha\sin\beta}{\sqrt{2}}\sum\limits_{j=0,1}\left\vert E_2\right\rangle_j$, in which $\vert gs\rangle$, $\vert E_1\rangle_i$ and $\vert E_2\rangle_j$
are the eigenstates of the ground state, the first and second excited states respectively.
The two parameters $\alpha$, $\beta$  can be related to the rotational parameters in the circuit.

\vspace{10pt}

\section*{\uppercase\expandafter{\romannumeral4}. Computing Eigenstates}

In this work, the quantum simulations  are performed on the open platform \textsl{Qiskit}~\cite{qiskit}.
The practical quantum computations are performed on the superconducting quantum processor $\textsl{IBM\_oslo}$,
which has 7 qubits. Its median {\cn} error is about 8.3e${-3}$ and its median readout error is
about 2.2e${-2}$.
 In addition to the number of  {\cn} gates,  the structure of the circuit and the architecture of quantum processor could also affect the accuracy.
  The transpiler can  optimize the executing circuit according
to the hardware architecture.
The calculations used 15000 shots for each measurements.

\subsection*{A. $\Omega=3$}

Firstly the ground state of $\Omega=3$ with 4 particles is solved by the quantum simulator.
The eigenspace is 3 dimensional as shown in Table I.
There are two variational parameters $\eta, \theta$.
The expectation value of the Hamiltonian given by the quantum simulator is -4.013, while the
the exact value if -4.0.
The resulted  variational parameters are $\theta_0=1.62\pi,\eta=1.26\pi$. The wave function of the ground state is $\vert gs\rangle_{sim}=0.588\left\vert\uparrow\uparrow\downarrow\right\rangle+0.554\left\vert\uparrow\downarrow\uparrow\right\rangle+0.590\left\vert\downarrow\uparrow\uparrow\right\rangle$, which is quite close to the exact wave function $0.577(\left\vert\uparrow\uparrow\downarrow\right\rangle+\left\vert\uparrow\downarrow\uparrow\right\rangle+\left\vert\downarrow\uparrow\uparrow\right\rangle)$, as their overlap is $0.9994$.

The excited states are simulated with the same circuit as for the ground state.
After the wave function of the ground state are obtained, the effective Hamiltonian
of excited states are constructed according to Eq.\ref{excit}.
Then excited states are solved by VQD with two variational parameters.
Note that the first excited state have a double degeneracy.
The two solutions correspond to parameters as as $\{1.8\pi,1.76\pi\}$ and $\{1.4\pi,0.42\pi\}$, respectively.
The resulted excited energies are -1.024 and -1.004 respectively, while the exact value is -1.0.
Note that wave functions of degenerate states are not uniquely determined.
The degenerate states are solved successively and the second degenerate state is obtained
by VQD after the first state is shifted out of the eigenspace via Eq.\ref{excit}.
The large deviations of higher states can be traced back to the accumulation of small deviations of lower states according to Eq.\ref{excit}.
Besides, the statistical uncertainties of quantum simulations also play a role.

Quantum computing of ground state and excited states of $\Omega=3$ are also displayed in Table II.
To compare with the simulation results, the fixed variational parameters are used in quantum computing.
For the ground state, the obtained energy is -3.75 with $\textsl{IBM\_oslo}$, which has a deviation due to the noisy hardware about 6\% compared to simulations.
The degenerate energies of  the first excited states are -1.13 and -1.11, respectively. The quantum-noisy deviations
of excited states are about 13\% with comparison with simulations. The deviations are mainly come from the noisy  {\cn} gates and the decoherence in readout measurements.

\subsection*{B. Error Mitigation}

Next we applied error mitigation of readout measurements of eigenstates of $\Omega=3$ with $\textsl{IBM\_oslo}$.
The detailed results are listed in Table II. 
For each qubit, there is a readout error probability of $P(0|1)$ and $P(1|0)$. The measurement
error can be mitigated by applying the inverse of the error matrix of $k$-th qubit~\cite{papenbrock,6Li}
\begin{equation}
S_k=
\displaystyle{\left(
\begin{array}{cc}
P_{0,0}^k & P_{0,1}^k \\
P_{1,0}^k & P_{1,1}^k \\
\end{array}
\right)}
\end{equation}
Note that such readout error mitigation is performed for individual qubits.
 The readout error mitigation
is demonstrated to be very useful to improve the quantum computing accuracy with a large
number of qubits~\cite{6Li}.
In our case, the ground state energy after the error mitigation is -3.871, while
the raw value is -3.751.
The error-mitigated first excited state energies are -1.107 and -1.086,
while the raw values are -1.126 and -1.109.
We see that the readout error mitigation is significant for the ground state
but has minor influences for excited states.

We also applied the zero-noise extrapolation~\cite{papenbrock} for the error mitigation of
 {\cn} gates.
For each measurement on $\textsl{IBM\_oslo}$, we add 2 and 4 additional  {\cn} gates.
Note that the product of two  {\cn} gates is the identity.
Based on these results, corresponding to 1, 3 and 5  {\cn} repetitions, the polynomial fitting can extrapolate
the error mitigated results with zero  {\cn} gate.
 This procedure of error mitigation has been widely adopted~\cite{cloud,papenbrock,lipkin1,stetcu}.
The higher-order extrapolation with more  {\cn} repetitions could be  too noisy to be helpful.
The linear fit is applied to get the error-mitigate results.
The ground state energy is -3.312 and -2.663 for 2 and 4 additional {\cn} gates.
The zero-noise extrapolation is -4.188 with 4  {\cn} gates and -4.04 with 2  {\cn} gates.
It seems that the extrapolation with 4  {\cn} gates is already too noisy.
Indeed, the output eigenvectors with  4 additional  {\cn} gates have serious decoherence.
The zero-noise extrapolation for the degenerated excited states with 2 additional {\cn} gates are
-1.064 and -0.978 respectively. The zero noise extrapolation method is a phenomenological error mitigation
but can improve the accuracy remarkably.

{\centering \subsection*{C. $\Omega=4$}
}

The problem of $\Omega=4$ is much more complex than that of $\Omega=3$.
For $\Omega=4$ with 4 particles, the eigenspace is 6 dimensional and 5 variational parameters are needed.
In this case, the circuit becomes complicated because we have to realize the entanglement of 6 basis on 4 qubits.
With 5 variational parameters, we have to employ classical optimization method to find
the global minima. We have applied the Nelder-Mead method which is a widely used
derivative-free optimization method for finding multi-dimensional global minima~\cite{NM}.
The Nelder-Mead method is based on the transformation of multi-dimensional simplex.
Note that the hybrid quantum-classical computation as a promising direction has been extensively studied~\cite{hybrid,6Li,lipkin2,klco},
since quantum computing is only superior on specific tasks.
The simulations with $\textsl{Qiskit}$  result in an energy of -5.958 for the ground state, while
the exact value is -6.0. Here the deviation is originated from statistical fluctuations and the multi-dimensional optimization.
Similar to $\Omega=3$, the excited states are solved successively by VQD.
The calculated energies of the first degenerate excited state are -2.020, -2.000, -2.004, respectively, while the exact values are -2.0.
The energies of the second degenerate excited state are -0.064 and -0.015, respectively, while
the exact values are -0.0.
It can be seen that accumulated deviations increase for higher states.
The overlaps between eigen states are also checked.
The orthogonality is satisfactory and this is crucial for simulations of thermal states.

For quantum computing of $\Omega=4$ eigen states, the ground state is maximally entangled and
the result of the complex circuit is -4.534. Its deviation is about 24\%. The detailed results are shown in Table III. 
The quantum computing of the first degenerate excited states are rather satisfactory. 
However, the results of the second degenerate excited states have large deviations, since
more computing basis are entangled compared to the first excited states. 
Note that
in the $\textsl{IBM\_oslo}$ processor some qubits are not directly connected.
The non-local operations can results in large noise.
For $\Omega=4$, the readout error mitigation and zero-noise extrapolation have also been performed.
We see that results can be much improved by the zero-noise extrapolation. 
The largest deviation for $\Omega=4$ after the zero-noise extrapolation is about 0.6, which is much larger than  0.1 for $\Omega=3$. 
The circuit in Fig.2 works for both ground state and excited states of $\Omega=4$.
For testing calculations, we also construct another circuit only for the first excited state on
4 qubits. The first excited state only involve the superposition of two basis as shown in Table I,
so that the circuit is simpler. The resulted energy with $\textsl{IBM\_oslo}$
is $-1.97\pm 0.03$,while the $\textsl{Qiskit}$ value is -1.993.  We see that the reduced eigenspace
can improve the accuracy.
In Ref.~\cite{6Li}, the first excited state of $^6$Li also employs a simpler circuit and has a more accurate result compared to
 the ground state considering their different eigenspaces.

\begin{figure}[htb]
\begin{center}
\includegraphics[width=0.42\textwidth]{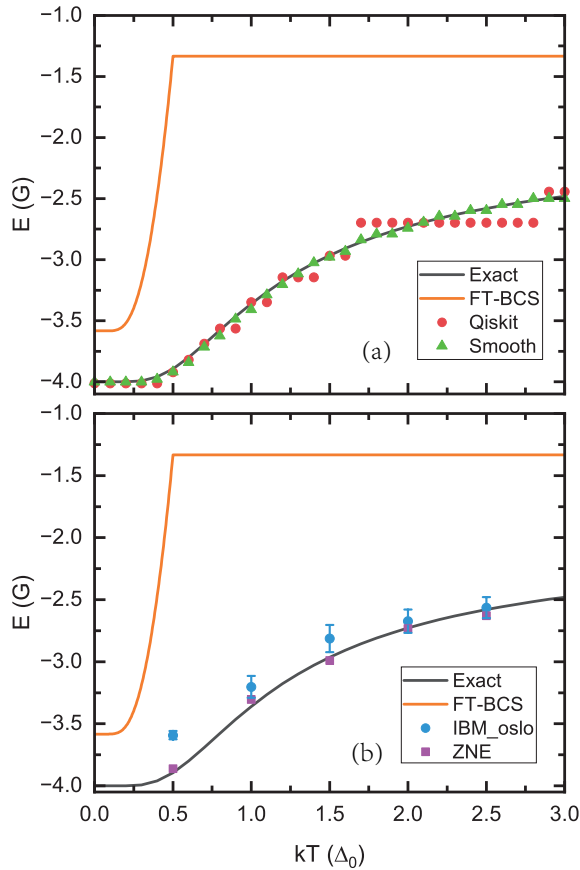}
\caption{
(a) The pairing energies of $\Omega=3$ as a function of temperatures obtained from FT-BCS and $\textsl{Qiskit}$.
The original $\textsl{Qiskit}$ and smoothed simulation results are shown. 
(b) The same pairing energies and zero-noise extrapolation (ZNE) results obtained by quantum computing.
}
\label{FIG3}
\end{center}
\end{figure}

\begin{figure}[tb]
\begin{center}
\includegraphics[width=0.42\textwidth]{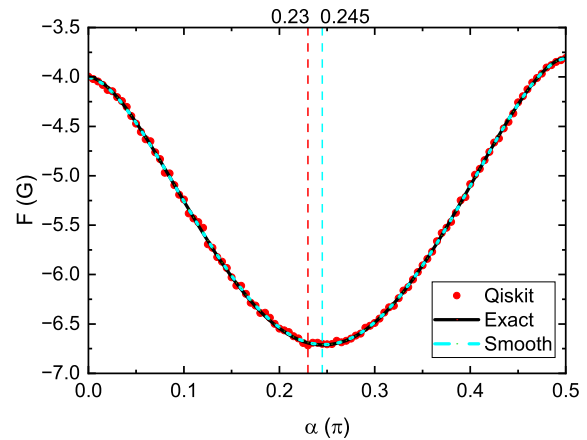}
\caption{
The simulated free energies of $\Omega=3$ as  a function of variational parameters at the temperature of $kT=2.7\Delta_0$.
The optimized variational parameters from $\textsl{Qiskit}$ simulations and smoothed simulations are shown for comparison.
}
\label{FIG1}
\end{center}
\end{figure}

\vspace{10pt}

\section*{\uppercase\expandafter{\romannumeral5}. Results at Finite Temperatures}

The thermal excitation states can be constructed with the eigenstates provided by zero-temperature calculations.
These eigenstates are almost completely orthogonal so that an entanglement state can approximate the mixed thermal state.
The exact solutions are given by the Seniority model.
For comparison, the FT-BCS results which are not suitable for finite systems are also shown.

{\centering \subsection*{A. $\Omega=3$}
}

For $\Omega=3$ with $N$=4 particles, the $\textsl{Qiskit}$ simulations of pairing energies as a function of temperatures are shown in Fig.3.
The temperature is given in the scale of $\Delta_0$.
With the FT-BCS approximation, there is a phase transition around temperature $kT$=0.5$\Delta_0$ as expected, which should be smoothed
out in finite systems.
The ground state energy of BCS is -3.55.
At high temperatures, the FT-BCS energy is -1.33 which is contributed from the Hartree term and pairing gap is vanished.
We see there is a significant discrepancy between FT-BCS and exact results.
The exact energies show a smooth transition as a function of temperatures.
The FT-BCS results are higher than  exact results both at zero and finite temperatures,
since BCS includes insufficient many-body correlations.
With increasing temperatures, the pairs are breaking due to thermal excitations.
However, the pairs can not fully broken due to a restricted configuration space.
At the limit of high temperatures, the system would have the largest entropy and eigenstates are equally mixed,
and the energy limit should be -2.0 rather than -1.33 given by FT-BCS.

The false phase transition from FT-BCS demonstrated the breakdown of the BCS approximation for small systems.
Note that BCS violates the conservation of particle numbers due to the breaking of U(1) symmetry.
The symmetry restoration by particle number projection can improve the FT-BCS results~\cite{Esebbag}.
However, the results of projected FT-BCS at high temperatures are close to FT-BCS~\cite{Esebbag},
which is not consistent with exact results of the seniority model.
It is known that FT-BCS with  variation after projection can well reproduce the non-degenerate pairing model~\cite{lacrox3}.
The energy discrepancy between exact solutions and FT-BCS at high temperatures demonstrates an analogy existence of a pseudogap pairing~\cite{pseudo},
i.e., a gap above $T_c$ is needed in Eq.(\ref{ftbcs}) to account for the energy discrepancy.
Here the existence of the psedogap pairing at high temperatures is due to the symmetry constraint of finite systems.
This provides a clue for the origin of the pseudogap phase in high-$T_c$ superconductors which may be induced by
specific localized symmetries.
The breakdown of FT-BCS and projected FT-BCS implies
the accurate treatment of many-body correlations in such small thermal-excited systems is essential.

\begin{figure}[tb]
\begin{center}
\includegraphics[width=0.227\textwidth]{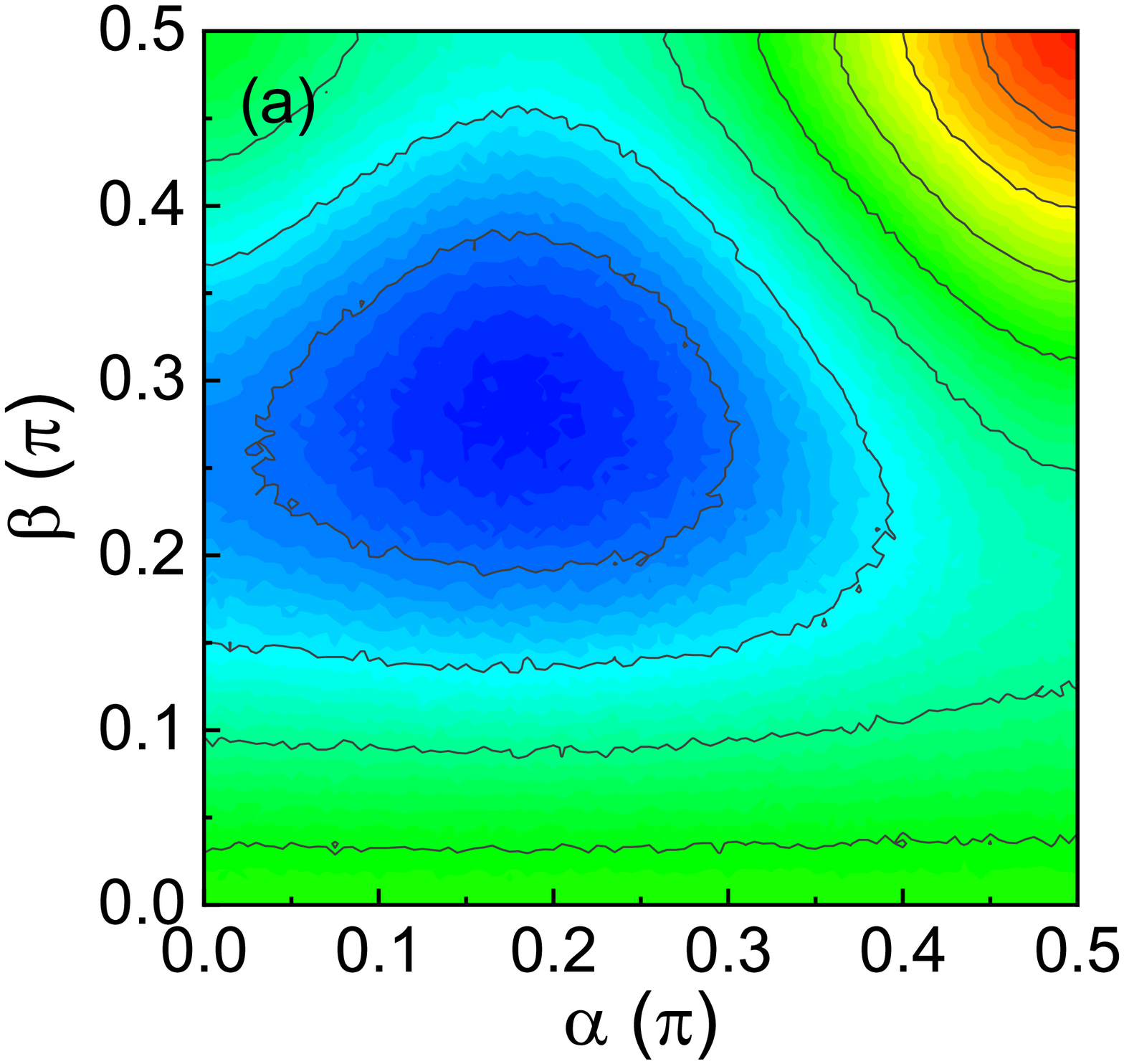}
\includegraphics[width=0.245\textwidth]{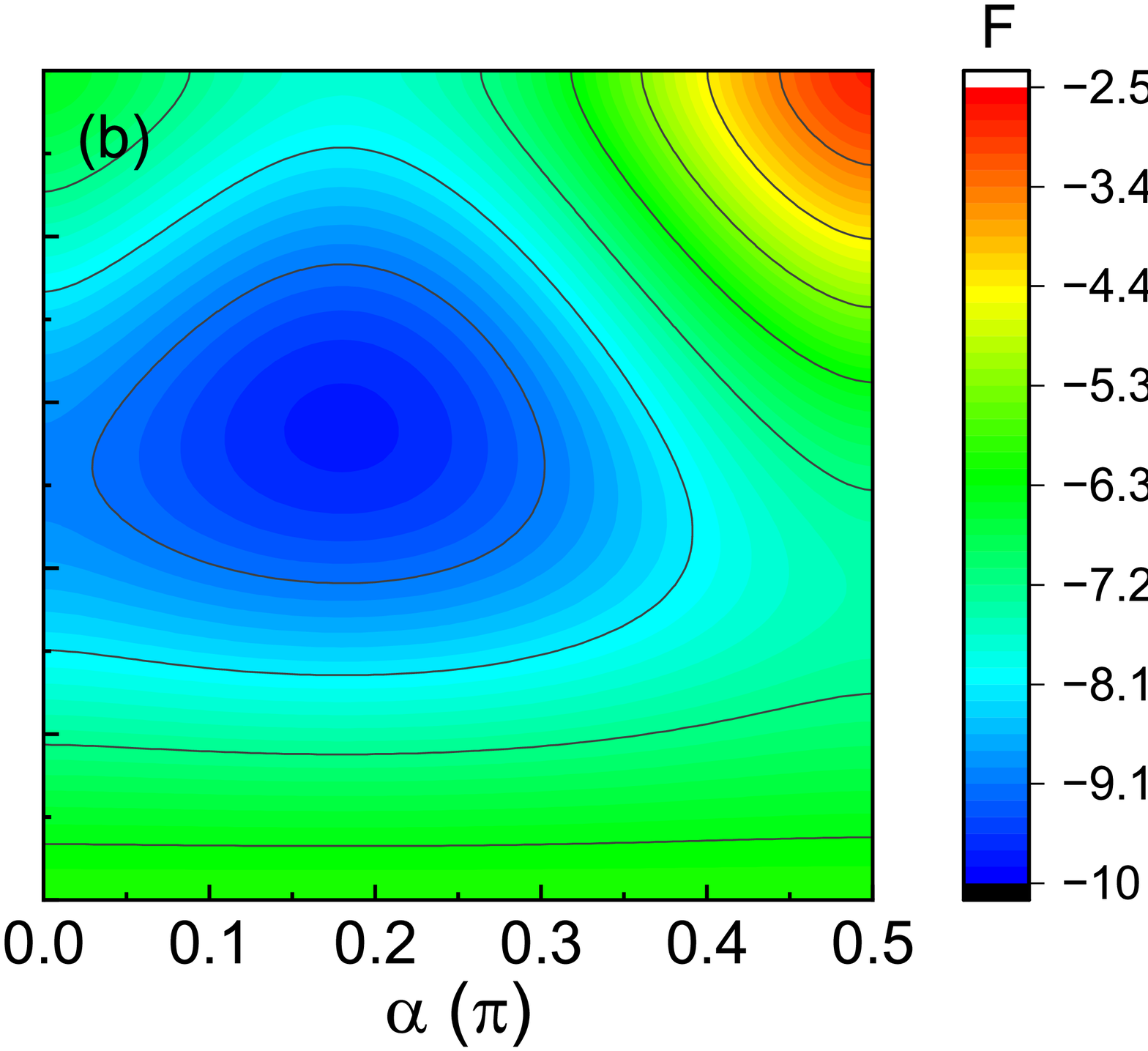}
\caption{
(a) The contour of free energies with two variational parameters by quantum simulations at $kT=2.0\Delta_0$ for $\Omega=4$.
(b) the smoothed contour around the minima.
}
\label{FIG1}
\end{center}
\end{figure}

With $\textsl{Qiskit}$ simulations, the deviations caused by statistical fluctuations are much larger at finite temperatures
than at zero temperature. This is because the calculation of entropy based on wave function overlaps adds more statistical noises.
In Fig.3, to reduce the noise, only one variational parameter is used considering the degeneracy of the first excited state.
The original $\textsl{Qiskit}$ simulations show large deviations from exact values.
To this end, we applied numerical smoothing and interpolation to smooth out the statistical fluctuations.
Then the smoothed results are close to exact values at finite temperatures.
As an example, the variational free energy at $kT$=2.7$\Delta_0$ is shown in Fig.4.
We see that the $\textsl{Qiskit}$  simulations have small fluctuations around exact free energies.
This results in small deviations in determining the variational parameter.
However, the determination of pairing energies of thermal states is sensitive to these variational parameters.
Thus the smoothed VQE is necessary to improve the accuracies of thermal quantum simulations.

The quantum computing results with $\textsl{IBM\_oslo}$ are shown in Fig.3(b).
Note that in the quantum computing, the variational parameters are fixed to reduce uncertainties.
We can see that the quantum computing results are slightly higher than the exact values and
also demonstrate the smooth phase transition.
For each temperature, we made 5 measurements and the hardware uncertainties are also shown.
The earlier described error mitigation by zero-noise extrapolation is also applied at finite temperatures.
The error mitigated results become close to exact values.

\begin{figure}[tb]
\begin{center}
\includegraphics[width=0.45\textwidth]{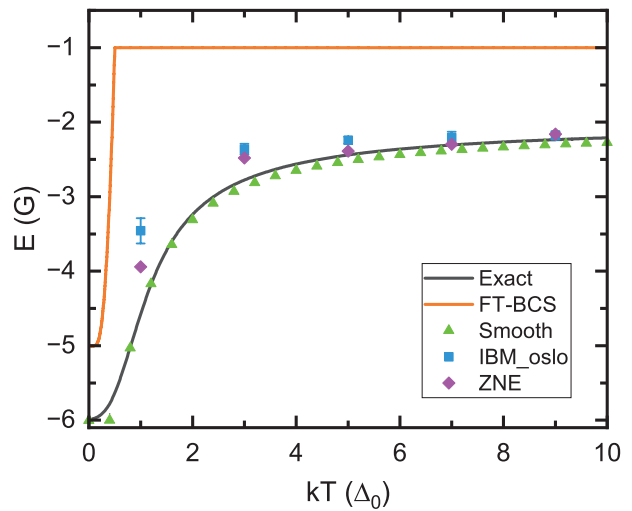}
\caption{
The pairing energies and the zero-noise extrapolation (ZNE) from quantum computing as a function of temperatures for  $\Omega=4$.
The FT-BCS results and exact results are shown for comparison.
The smoothed simulation results by $\textsl{Qiskit}$ are also shown.
}
\label{FIG6}
\end{center}
\end{figure}

{\centering \subsection*{B. $\Omega=4$}
}

For $\Omega=4$, the quantum simulations at finite temperatures are more complex.
We adopt two variational parameters considering the degeneracy of excited states.
In principle there are 5 variational parameters but the influences of statistical noise would be very large.
Fig.5 shows the $\textsl{Qiskit}$ simulations of free energies with two variational parameters.
We see notable fluctuations in the contour which would be difficult for VQE to determine precisely the variational parameters.
Fig.5(b) shows the smoothed contour with the Fourier expansion.
The smoothed VQE can also be applied to the multi-dimensional parameter space.

The thermal excitation energies of the pairing Hamiltonian of $\Omega=4$ at finite temperatures are shown in Fig.6.
The FT-BCS results are also shown.
It is known that the original $\textsl{Qiskit}$ results have notable fluctuations in free energies.
The smoothed simulations are necessary to obtain correct parameters and thus correct pairing energies. 
The quantum computing results and zero-noise extrapolation are shown. 
The zero-noise extrapolation results are satisfactory compared to exact solutions. 
In general, the quantum  computing results of $\Omega=4$ are less accurate compared to that of $\Omega=3$.
In both cases, the deviations of quantum computing become smallers with increasing temperatures, due to the cancellation between errors of different states. 
This is promising for quantum computing of thermal states although accumulated errors by VQD increase at higher states. 
For even larger systems or very high temperatures, a hybrid approach can be adopted, in which
low-lying states are computed by VQD and high-lying states are computed by approximated methods
such as the mean-field approximation with symmetry projections.

\vspace{10pt}

\section*{\uppercase\expandafter{\romannumeral6}. Conclusion}

We performed quantum computing of eigenstates and thermal states of the pairing Hamiltonian in a degenerate shell.
For $\Omega=3$ and $\Omega=4$ with 4 particles, we show their wave functions can be simulated on 3 and 4 qubits,
which correspond to much larger shell model spaces.
We have applied VQD for excited states that shifts
lower states out of the eigenspace successively. The quantum computing is performed
with a superconducting quantum processor provided by IBM.
The error mitigation of readout measurements and the zero-noise extrapolation have been demonstrated to be helpful to improve
the accuracy.
 For  $\Omega=4$, the entanglement of 6 basis on 4 qubits is realized
 and the circuit is constructed using the two-qubit building blocks that preserves the number of particles.

 The mixed thermal state is simulated by the entanglement of the orthogonal eigenspace
 with the same variational circuit as at zero temperature.
 For comparison, the finite-temperature BCS results are also shown, which has
 a false phase transition from superfluid state to normal state. The quantum
 computing demonstrates a smooth transition as expected in finite systems.
 The FT-BCS  is breakdown for small systems.
In addition, exact results are not close to FT-BCS at high temperatures, indicating an analogy existence of pairing pseudogap.
In our approach, the thermal excitations can be simulated  without accurate
knowledge of eigenvalues. 
The results from quantum computing become close to exact solutions
at high temperatures.
In the future, it is still desirable to develop an improved quantum algorithm to compute the entropy. 
The accurate treatment of  many-body correlations
 of finite thermal systems has broad physics implications, for which quantum computing has unique opportunities.

\vspace{10pt}

\acknowledgments
We are grateful to discussions with F.R. Xu. 
This work was supported by the National Key R$\&$D Program of China (Contract No. 2018YFA0404403)
and by the National Natural Science Foundation of China under Grants
No.  11975032, 11835001, 11790325, and 11961141003. We acknowledge the use of IBM Q cloud and the Qiskit software package.

\end{document}